\documentclass[    ,final      ]{aipproc}
\layoutstyle{6x9}
\usepackage{amsfonts,amsmath,amssymb,mathrsfs}
\begin{document}

\title{The Message of the Quantum?}

\classification{
  03.65.Ta. 
  }
\keywords{
  Reality and information; no-hidden-variables theorems; 
  determinism and quantum mechanics. 
  }
\author{Martin Daumer}{address={Sylvia Lawry Centre for 
    Multiple Sclerosis Research, Hohenlindenerstr.\ 1, 
    81677 M\"unchen, Germany.\\
    e-mail: daumer@slcmsr.org.}}
\author{Detlef D\"urr}{address={Mathematisches Institut,
    Ludwig-Maximilians-Universit\"at, Theresienstr.\ 39, 80333
    M\"unchen, Germany.\\ e-mail: duerr@mathematik.uni-muenchen.de}}
\author{Sheldon Goldstein}{address={Departments of Mathematics and Physics,
     Hill Center, Rutgers, The State University of New
     Jersey, 110 Frelinghuysen Road, Piscataway, NJ 08854-8019, USA.\\
     e-mail: oldstein@math.rutgers.edu}}
\author{Tim Maudlin}{address={Department of Philosophy, Davison Hall,
     Rutgers, The State University of New Jersey, 26 Nichol Avenue,
     New Brunswick, NJ 08901-1411, USA.\\
     e-mail: maudlin@rci.rutgers.edu}}
\author{Roderich Tumulka}{address={Mathematisches Institut,
    Eberhard-Karls-Universit\"at, Auf der Morgenstelle 10, 72076
    T\"ubingen, Germany.\\ e-mail:
    tumulka@everest.mathematik.uni-tuebingen.de}}
\author{Nino Zangh\`\i}{address={Dipartimento di Fisica dell'Universit\`a di
    Genova and INFN sezione di Genova, Via Dodecaneso 33, 16146
    Genova, Italy.\\ e-mail: zanghi@ge.infn.it}}

\begin{abstract}
We criticize speculations to the effect that quantum mechanics is fundamentally about information. We do this by pointing out how unfounded such speculations in fact are. Our analysis focuses on the dubious claims of this kind recently made by Anton Zeilinger.
\end{abstract}

\maketitle

Quantum theory has always invited rather extreme speculations about the
nature of physical reality. John Wheeler \cite{wheeler}, for example, famously
conjectured that quantum mechanics suggests a ``participatory universe'' in
which the present observations of experimentalists can give ``tangible
reality'' to the distant past, that current actions can somehow
\textit{produce} the past physical structure of the universe, rather than
merely \textit{inform} us about it. It is a bold speculation, but one
underpinned by nothing in quantum mechanics itself. Neither the
experimental consequences of quantum mechanics nor any presently existing
precise understanding of the theory support this astonishing
suggestion. ``Quantum mechanics'' cannot justify this speculation because
precisely formulated theories that recover the quantum mechanical
predictions do not posit any such backward effect. So one sure check on
claims about ``the lesson of quantum theory'' can be obtained by considering
precise theories that recover all of the quantum predictions that have
ever been verified. Several such theories exist.

Wheeler's rather obscure ideas appear to live on today in the remarkable suggestion that physics is only about information or, even more astonishingly, that the physical world itself just \emph{is} information. The first suggestion would seem to contradict the everyday belief that physics is concerned with the physical structure of objects and the laws governing that structure: it is, therefore, about molecules and atoms and stars and electrons, among other things. Electrons can be \emph{used} in systems that convey information, as in telephone lines, but that is a rather specialized set of circumstances, and physics should cover all of the universe, not just special systems. As for the suggestion that the physical world just  \emph{is} information, the suggestion sounds more mystical than scientific. If it were to be put forward as a serious proposal it would need both clarification and powerful justification. Unfortunately, when the topic of discussion is quantum theory, basic standards of clarity and argumentation seem to be abandoned, even in the most prestigious journals.

A conspicuous example   has recently been published by Anton Zeilinger. He has put forward, in his essay ``The message of the
quantum'' \cite{Zei05}, some thoughts about the relationship between
reality and information, very much in the spirit of the traditional
``Copenhagen'' view of quantum mechanics. He has accompanied his opinion
with his personal summary of the conclusions that one should draw from
quantum mechanics at the centennial of Einstein's annus
mirabilis. Unfortunately, his claims are at best dubious, and most of them
are simply wrong.

Zeilinger writes that ``The discovery that individual events are
irreducibly random is probably one of the most significant findings of the
twentieth century.'' He claims, in other words, that determinism has been
refuted, that it has been proven that (some) individual events in the
quantum world are irreducibly random, rather than merely seeming random
because of our ignorance. This conclusion has been challenged, most
famously by Einstein. On what basis does Zeilinger conclude that Einstein
was wrong?  He presumably relies on the various no-go or
no-hidden-variables theorems---of von Neumann, Bell, Kochen and Specker, and
the like---which are supposed to show that quantum randomness can't be
regarded as arising merely from ignorance.  However, in his seminal paper
\cite{Bell87b} ``On the problem of hidden variables in quantum mechanics,''
John Bell has shown that these theorems involve unwarranted assumptions,
and thus don't justify the rejection of Einstein's view about the origin of
quantum randomness.

Perhaps Zeilinger merely means that the predictions of quantum theory---or
at least the experimental facts on which quantum mechanics is
based---strongly support a non-deterministic formulation. But this view is
easily refuted by the counter-example provided by Bohmian mechanics
\cite{Bohm52,Gol01}, a theory describing the deterministic evolution of
particles that accounts for all of Zeilinger's examples and indeed all of the
phenomena of nonrelativistic quantum mechanics, from spectral lines to the
two-slit experiment and random decay times. Thus, the experimental facts of
quantum mechanics do not establish indeterminism. At best, which
explanation of the experimental facts to prefer, Bohm's simple
deterministic one or the convoluted indeterministic one of the Copenhagen
view, remains our theoretical choice.

 Zeilinger suggests that ``one could find comfort'' in the idea of
determinism, if it were tenable. This suggestion, whatever its merits,
gives entirely the wrong impression of the main motivation of the critics
of the Copenhagen view. David Bohm and John Bell \cite{Bell87b}, two of its
leading critics, did not hesitate to use stochastic theories---with
irreducible randomness---when that served a purpose. Even Einstein, often
inappropriately depicted as a stubborn adherent of determinism claiming
that ``God does not play dice,'' made it clear in his ``Reply to Critics''
\cite{Ein49} that he rejected the Copenhagen view not because of its
indeterminism but because it fails to describe quantum phenomena in terms
of objective events occurring independently of subjective perceptions.

Zeilinger writes that ``John Bell showed that the quantum predictions for
entanglement are in conflict with local realism.'' In fact, realism was not
among the assumptions Bell used for deriving the conflict with quantum
mechanics, even though realism about spin observables---which is much more
than realism in general---occurred in the argument as an implication  of
locality. What Bell proved is that the predictions of quantum theory for
spin correlations are incompatible with locality, i.e., that quantum
mechanics is irreducibly \emph{nonlocal}, a point that Bell \cite{Bell87b} repeatedly stressed.

In Zeilinger's view, Bell's result suggests, not that there is nonlocality,
but that ``the concept of reality itself is at stake.''  What that is
supposed to mean is left vague, but it is hard to see what the meaning
could be that does not ultimately lead to the view that nothing exists
objectively outside our minds. That is not a scientific view and there is
nothing in Bell's result to support it. And contrary to Zeilinger's claim,
it is not supported by the Kochen--Specker theorem either, as Bell was the
first to point out: It cannot be maintained that the Kochen-Specker paradox
supports the notion that there is a problem with ``the concept of reality
itself'' when there are perfectly realistic theories, such as Bohmian
mechanics or the Ghirardi--Rimini--Weber version of quantum mechanics
involving spontaneous random collapse of the wave function \cite{grw}, that account for all of the experimental facts on which that paradox is based.

Interestingly, Zeilinger draws the correct moral from the Kochen--Specker
paradox when he writes: ``even for single particles, it is not always
possible to assign definite measurement outcomes independently of and prior
to the selection of specific measurement apparatus in the specific
experiment.'' That is, in giving a physical account of a measurement, we
must take account of the exact physics of the experimental
situation. Indeed, Bohmian mechanics and the Ghirardi--Rimini--Weber version
of quantum mechanics allow us to do precisely this, since they do not
postulate some special physics for measurements. And once done, all the
predictions come out right.

Most baffling, Zeilinger suggests that ``the distinction between
reality and our knowledge of reality, between reality and
information, cannot be made.''  After such a counterintuitive
assertion, we naturally expect to find a mighty argument in its
behalf. Here, however, is Zeilinger's: ``There is no way to refer to
reality without using the information we have about it.'' In other
words, what we can say about reality, or better what we can know
about reality, must correspond to our information about reality. In
other words, what we know about reality must conform to what we know
about reality. Does Zeilinger really believe that a tautology such
as this can have interesting consequences?

 The very concepts of knowledge and information imply a special kind of
relationship between different things, appropriate correlations between a
knower and what is known. Thus ``the distinction between reality and our
knowledge of reality'' not only can be made; {\em it must be made if the
notions of knowledge and information are to have any meaning in the first
place.}

At a time when the forces of obfuscation in America are engaged in a
campaign against the theory of evolution on behalf of Intelligent Design,
it is perhaps worth asking Zeilinger how the idea that there is no
difference between information and reality can be compatible with the
emergence of information processing systems such as we are from a lifeless
reality. And it is perhaps also worth asking the editors of Nature how, at
a time when, rightly, papers on Intelligent Design are consistently
rejected by peer-reviewed journals, an essay like Zeilinger's is not.

\begin{theacknowledgments}
 M.~Daumer is supported in part by the German National Science Foundation 
 DFG, SFB 386. S.~Goldstein is supported in part by NSF Grant DMS-0504504.
 N.~Zangh\`\i\ is supported in part by INFN.
\end{theacknowledgments}

\bibliographystyle{aipproc}

\end{document}